\documentclass[letterpaper, 10 pt, conference]{ieeeconf}  

\usepackage{cite}
 
\IEEEoverridecommandlockouts                              

\usepackage{graphicx} 
\usepackage{amsmath} 
\usepackage{amssymb}  

\usepackage{hyperref}
\usepackage[normalem]{ulem}
\usepackage{siunitx}
\usepackage{stfloats}
\usepackage{bm}
\usepackage{multirow}
\usepackage{multicol}
\usepackage{hyperref}

\usepackage{fancyhdr}

\title{\LARGE \bf

End-to-End Low-Level Neural Control\\of an Industrial-Grade 6D Magnetic Levitation System

}

\author{Philipp Hartmann$^{1}$, Jannick Stranghöner$^{1}$, and Klaus Neumann$^{1,2}$%
}

\begin{document}
\raggedbottom
\bstctlcite{IEEEexample:BSTcontrol}
\maketitle
\thispagestyle{empty}
\pagestyle{empty}

\fancypagestyle{firstpage}{%
	\fancyhead{}
	\renewcommand\headrule{}
	\fancyfoot[C]{%
		\scriptsize 
		Accepted for publication at the 2026 IEEE International Conference on Robotics and Automation (ICRA).\\
		\copyright 2026 IEEE. Personal use of this material is permitted. Permission from IEEE must be obtained for all other uses, in any current or future media, including reprinting/republishing this material for advertising or promotional purposes, creating new collective works, for resale or redistribution to servers or lists, or reuse of any copyrighted component of this work in other works.%
	}
}
\thispagestyle{firstpage}



\begin{abstract}
Magnetic levitation is poised to revolutionize industrial automation by integrating flexible in-machine product transport and seamless manipulation. It is expected to become the standard drive technology for automated manufacturing. However, controlling such systems is inherently challenging due to their complex, unstable dynamics. Traditional control approaches, which rely on hand-crafted control engineering, typically yield robust but conservative solutions, with their performance closely tied to the expertise of the engineering team. In contrast, learning-based neural control presents a promising alternative. 
This paper presents the first neural controller for 6D magnetic levitation. Trained end-to-end on interaction data from a proprietary controller, it directly maps raw sensor data and 6D reference poses to coil current commands. The neural controller can effectively generalize to previously unseen situations while maintaining accurate and robust control. 
These results underscore the practical feasibility of learning-based neural control in complex physical systems and suggest a future where such a paradigm could enhance or even substitute traditional engineering approaches in demanding real-world applications.
The trained neural controller, source code, and demonstration videos are publicly available at \mbox{\url{https://sites.google.com/view/neural-maglev}}.

\end{abstract}
\section{Introduction}
\setcounter{footnote}{1}
\footnotetext[1]{CITEC, Faculty of Technology, Bielefeld University, Germany.}
\footnotetext[2]{Fraunhofer IOSB-INA, Lemgo, Germany.}

\begingroup
  \renewcommand{\thefootnote}{}%
  \footnotetext{Correspondence to: \texttt{philipp.hartmann@uni-bielefeld.de}}%
  \addtocounter{footnote}{-1}%
\endgroup

\begingroup
  \renewcommand{\thefootnote}{}%
  \footnotetext{This work was supported by the Fraunhofer Internal Programs under Grant No. SME 40-09551.}%
  \addtocounter{footnote}{-1}%
\endgroup

Magnetic levitation~(MagLev) with six degrees of freedom~(6D) is, by any engineering standard, an absolute technology highlight. The ability to transport objects without friction on movers that can fly freely over planar surfaces, decoupled from mechanical wear and abrasion, opens new directions in packaging, assembly, and factory automation~\cite{beckhoff2025pccontrol},~\cite{ wreeManufacturingSystemsIndividualized2025}. Consequently, several major companies in the automation industry have integrated magnetic levitation systems into their product portfolios~\cite{sandhofnerACOPOS6DBeginn2021, schultePlanarmotorantriebssystemXPlanar2022b, PlanarMotor_nodate, BoschRexroth2022_FLOW6D}. Yet, this remarkable flexibility comes at a price: MagLev systems are inherently difficult to control at the lower level, where high-frequency and precise stabilization of the mover in 6D is required. This difficulty is a consequence of Earnshaw's theorem~\cite{griffithsIntroductionElectrodynamics2013}, which precludes stable levitation in static magnetic fields. As a result, even small modeling errors or unmodeled effects can lead to a rapidly growing instability.

\begin{figure}[thpb]
      
      \centering
      \includegraphics[width=\columnwidth]{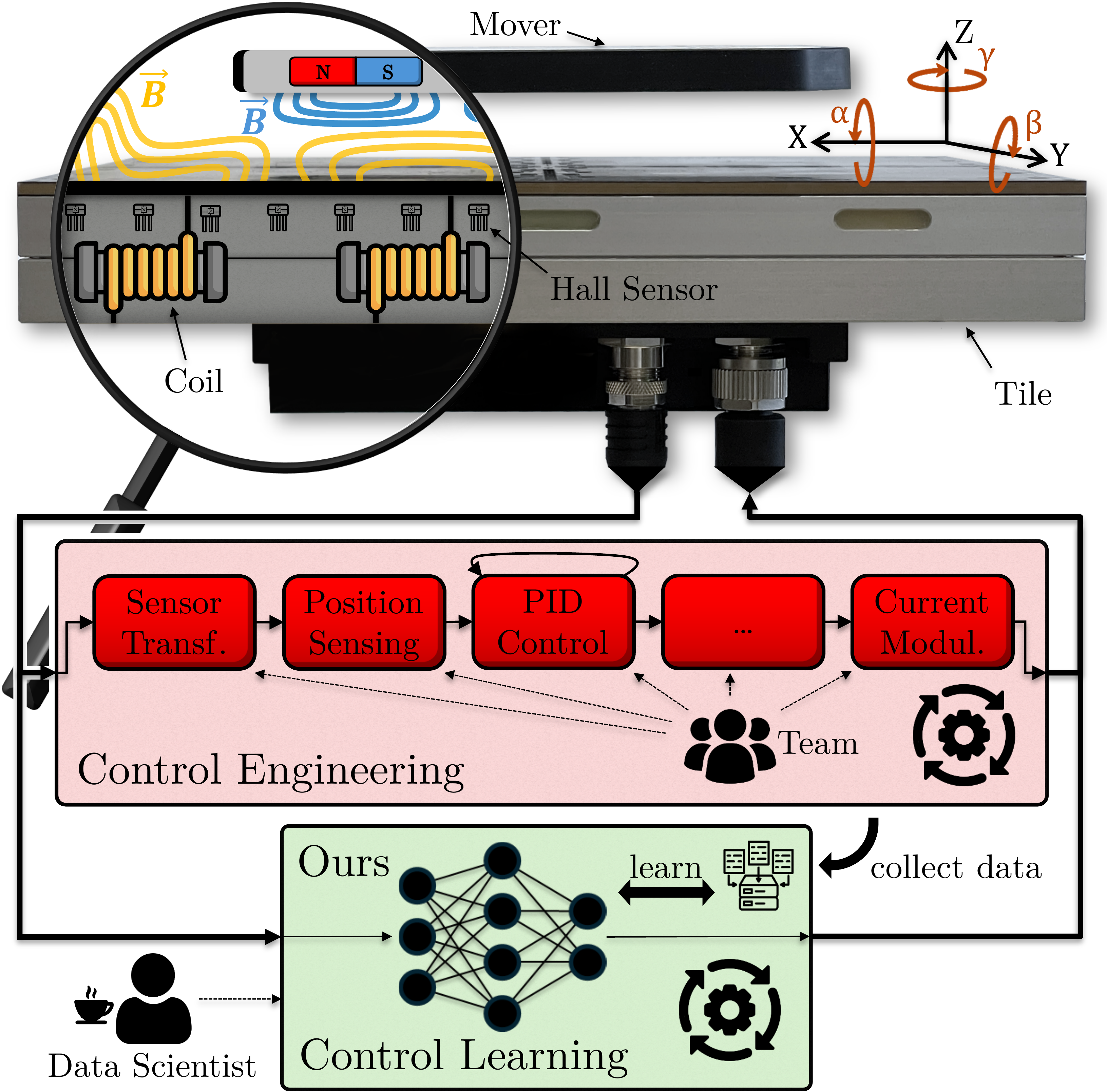}
      \vspace{-1.6em} 
      \caption{Traditional engineering approaches (red box) involve expert teams that hand-craft control pipelines typically consisting of numerous modules based on a system model. However, 6D magnetic levitation is complex and therefore, model assumptions may differ from the real system -- known as model mismatch. Our neural control approach (green box) can inherently capture all systematics by learning from interaction data from the system in an \mbox{end-to-end} fashion.}
      \vspace{-1.5em} 
      \label{fig:VA}
\end{figure}

Therefore, the performance of MagLev controllers is often closely linked to the expertise and resources of the engineering team, as illustrated in Fig.~\ref{fig:VA}. Modern pipelines rely on domain knowledge to explicitly model disturbances such as temperature drift, actuator cross-coupling, and manufacturing tolerances, but certain effects are inherently difficult to capture, either due to a lack of data, excessive modeling complexity, or incorrect assumptions made by the expert team. These “blind spots” often manifest only during deployment, when disturbances combine in unexpected ways and lead to degraded performance or even instability due to model mismatch. In this context, each incremental improvement during development requires additional modularization, calibration, and human intervention. 
Consequently, achieving high-precision levitation requires multi-year development cycles, where costly expert teams must manually design and tune the control pipeline.

However, motion performance is a critical determinant of manufacturing quality and productivity~\cite{zhouMagneticLevitationTechnology2022a}, enforcing narrow margins for modeling error. In contrast, learning-based neural approaches have recently been demonstrated to be a powerful alternative~\cite{brunkeSafeLearningRobotics2022, bergmannPrecisionFocusedReinforcementLearning2024, biekerDeepModelPredictive2020, Neumann2015}, particularly for systems where explicit modeling is either impractical or highly susceptible to such errors. These approaches have achieved remarkable success in a variety of complex domains by learning control strategies directly from interaction data~\cite{mnihHumanlevelControlDeep2015, tangDeepReinforcementLearning2025}. Therefore, inspired by Nobel laureate Demis Hassabis, who remarked "what we have always found is that the more end-to-end you can make it, the better the system"~\cite{Fridman2022Hassabis299}, a natural question arises: what if the system itself could serve as its own experimental platform, reducing modeling effort to a minimum through an end-to-end (E2E) approach? 
This would allow controllers to improve directly through interaction with the system, autonomously learning relevant dynamics, including complex disturbances, while seamlessly adapting to manufacturing tolerances and temporal drift. Additionally, this opens up the prospect of quantitative improvements, such as increased energy efficiency, robustness, and accuracy, as well as qualitatively new behaviors, such as an unconstrained \SI{360}{^\circ} rotation around the $z$-axis.
Viewed from this angle, the broader vision is to establish learning-based neural control as a new paradigm for control engineering. In the context of MagLev, this is an ambitious and novel direction. To date, practical MagLev control has been exclusively the domain of modular, manually engineered pipelines.

In this paper, we introduce the first E2E neural controller for a 6D MagLev system. We present this integration of hard real-time neural inference with an industrial magnetic levitation architecture as a novel system-level contribution. To enable this, we publish a highly optimized real-time-capable inference library for feedforward and recurrent neural networks. 
Next, we empirically analyze the accuracy and robustness of the neural controller, especially dealing with covariate shift by compensating for systematic pose deviations. 
Experiments with a real MagLev system confirm robust, accurate control, strong generalization, and even extrapolation to unseen reference poses and system dynamics. These results demonstrate the potential of E2E neural control in MagLev as a representative case of industrial automation, establishing an initial baseline for future research. 

\section{Related Work}
\label{sec:RelatedWork}
\begin{figure*}[t]
    \centering
    \includegraphics[width=\textwidth]{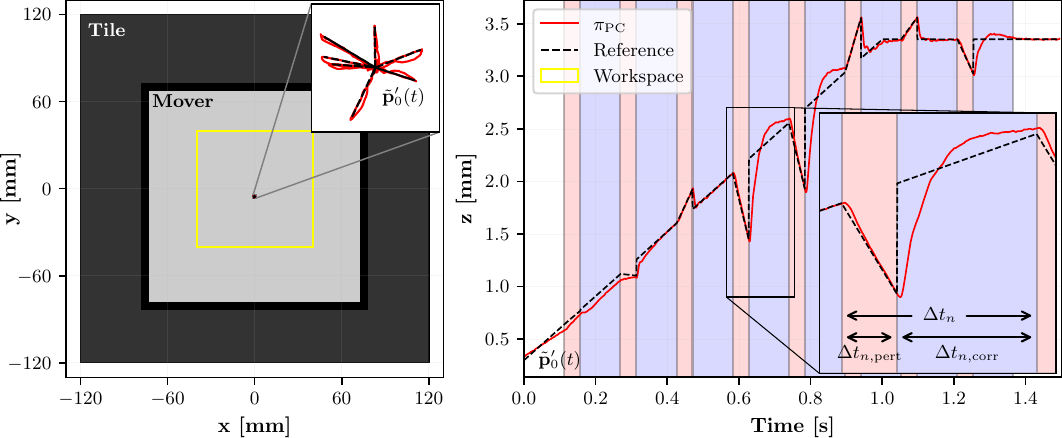}
    \vspace{-1.6em} 
    \caption{The defined workspace (yellow) illustrates all valid mover poses over a one-tile system in $x$ and $y$. Data is generated from responses of a proprietary controller to lift-off trajectories $\tilde{\mathbf{p}}'_{n}(t)$, each with constant, randomly sampled values for $x, y, \alpha, \beta, \gamma$. 
    Lift-off and stable hovering mainly affects the $z$-axis, shown on the right. To improve robustness, all dimensions are augmented by reference pose jumps, consisting of a perturbation phase (red, duration $\Delta t_{n,\mathrm{pert}}$) followed by a correction phase (blue, duration $\Delta t_{n,\mathrm{corr}}$).}
    \vspace{-1.5em}
    \label{fig:System}
\end{figure*}

MagLev systems have been developed with various hardware architectures~\cite{zhouMagneticLevitationTechnology2022a}. This work specifically addresses feedback-controlled MagLev, where permanent magnets on a passive motor module (mover) are localized by Hall sensors, enabling precise coil-driven levitation and 6D pose tracking.  MagLev systems are based exclusively on expert-engineered modular control pipelines~\cite{zhouMagneticLevitationTechnology2022a, zouForceTorqueModel2023a}, which require explicit calibration modules to mitigate actuator-induced interference~\cite{jansenMagneticallyLevitatedPlanar2008a},~\cite{zouForceTorqueModel2023a}. Such modular control pipelines typically incorporate coordinate transformations, sensor transformations, mover localizations, observers, PID controllers, force projections, current demodulation, coil switching, or amplifier stages~\cite{lu6DDirectdriveTechnology2012,lahdoDesignImplementationNew2019a, zouForceTorqueModel2023a, wangAdaptiveFixedTimePosition2023, xuDevelopmentMagneticallyLevitated2022}, while still emphasizing extensibility to compensate for disturbances such as temperature variation~\cite{lei6DegreesFreedom2023a}, geometric misalignment, and amplifier errors~\cite{nguyenErrorAnalysisMinimize2020a}. While many software modules have already been explored, there are still recent works focusing on specialized modules to improve the system's performance, such as comprehensive error analysis methods for cross-axis coupling minimization~\cite{xuDevelopmentMagneticallyLevitated2022}, large-angle 6D analytic wrench modeling via harmonic spectral field representation with Taylor-linearized rotational flux and dual‑frame decoupling~\cite{huDynamicModelingApproach2025}, or~\cite{xuPredictorbasedModelPredictive2023} explicit compensation of I/O and computation delays using predictor-based model predictive control. Although existing methods achieve accurate and robust control, they require substantial engineering expertise and iterative refinement, leading to brittle and complex solutions. 
While some recent works propose analytical formulations for position feedback~\cite{zhangLongRange6DOFAbsolute2024}, others employ neural network–based approaches embedded within modular frameworks demonstrating notable performance gains~\cite{schultePlanarmotorantriebssystemXPlanar2022b}. These results provide strong motivation for the exploration of fully E2E methodologies. 

E2E learning bypasses intermediate modules by directly mapping raw sensor input to actuator commands, simplifying controller design and deployment. Its effectiveness has been demonstrated in robotic manipulation, where E2E learning produces superior task-specific representations compared to modular control pipelines~\cite{levineEndtoEndTrainingDeep2016a}. Extending beyond robotic manipulation, this approach has been successfully applied in various control domains, including autonomous driving, where behavior cloning (BC) has demonstrated strong performance despite inherent challenges such as covariate shift~\cite{bojarskiEndEndLearning2016}. The covariate shift arises when the learned neural controller encounters unfamiliar states that are not present in the training data~\cite{zareSurveyImitationLearning2024b}. Strategies such as data augmentation, noise injection, and constrained data distributions are widely used to address these issues, improving robustness and generalization~\cite{popovMitigatingCovariateShift2024, laskeyDARTNoiseInjection2017a, bansalChauffeurNetLearningDrive2019, zareSurveyImitationLearning2024b}.

Although 6D MagLev systems lack prior work on neural control, their inherently unstable dynamics and high-frequency actuation share key characteristics with other domains. A representative example is low-level quadrotor flight control, where maintaining stability under changing dynamics demands precise actuation. Here, neural control based on model-based reinforcement learning replaced traditional PID control while improving robustness~\cite{lambertLowLevelControlQuadrotor2019, feredeEndtoendNeuralNetwork2024b}. Even in highly nonlinear systems such as a redundant musculoskeletal arm, neural control effectively handles dynamics that exceed the capabilities of traditional controllers~\cite{driessLearningControlRedundant2018a}. Similarly, neural approximations of model predictive control improve robustness and accuracy compared to PID control in robotic balancing tasks~\cite{hoseMiniWheelbotTestbed2025}. In high-frequency permanent magnet synchronous motor control domains, neural controllers have shown strong performance in directly predicting actuator commands~\cite{schenkeDeepQLearningDirect2021a, tomDesignMachineLearningbased2025}. Furthermore, deep reinforcement learning controllers trained in simulation have demonstrated adaptability and reduced manual tuning efforts for tokamak plasma shaping~\cite{degraveMagneticControlTokamak2022a}. 

In contrast to all previous work on 6D MagLev control, which relies on modular, expert-engineered controllers, this paper presents the first E2E neural controller for 6D MagLev. Our method directly maps raw Hall sensor measurements and 6D reference poses to coil-level actuation commands, entirely bypassing manual expert-engineered modules such as pose estimation. Although prior work has applied neural components within modular pipelines, no existing approach has learned closed-loop E2E neural control directly from interaction data for 6D MagLev. 
Although E2E control has been demonstrated in various related domains, such as quadrotor flight or plasma shaping, these systems operate at lower control frequencies, with simpler actuator structures, or demand simulation environments, limiting their direct applicability to MagLev. 
This work bridges the gap to industrial-grade, highly unstable domains operating at microsecond-level latencies.
Since no open-source baselines exist for 6D MagLev, we utilize the proprietary controller as the expert. By demonstrating BC on an industrial-grade system, this work establishes the first foundational benchmark for learning-based control in this domain.

\section{System Overview}
\label{sec:System_Overview}

The MagLev system used for evaluation is XPlanar~\cite{bentfeldMethodControllingPlanar2023, neumannMethodControllingPlanar2023, neumannMethodControllingPlanar2024} by Beckhoff Automation (see Fig.~\ref{fig:VA} and Fig.~\ref{fig:System}), operating in hard real-time at a control frequency of \SI{4}{\kilo\hertz} equal to a cycle time of \SI{250}{\micro\second}~\cite{wreeManufacturingSystemsIndividualized2025}. The existing proprietary controller with version number 4.2.11.0 is implemented in a dedicated control environment based on TwinCAT 3.1.4024.65 running on an industrial PC (C6030) equipped with an Intel i9-9900K CPU. 
All computation is restricted to CPU execution without hardware acceleration. The control logic operates within a single TwinCAT real-time task, restricted to one CPU core, and executed in kernel mode independently of the host operating system to ensure deterministic execution times. The active motor module (called \emph{tile}) measures $240\text{mm} \times 240\text{mm}$, containing 24 coils, 50 Hall effect sensors (HES), and 5 temperature sensors with product number APS4322-0000-0000. The passive motor module (called \emph{mover}) measures $155\text{mm} \times 155\text{mm}$, equipped with 4 Halbach arrays, each consisting of 5 permanent magnets with product number APM4330-0000-0000. This hardware configuration supports full 6D motion~\cite{beckhoff2025pccontrol}. This work utilizes one tile and one mover.

Low-level actuation is handled by the proprietary controller, acting as a black-box expert in our learning setup. At a higher level of abstraction, the proprietary controller $\pi_{\text{PC}}$ operating on this system can be expressed as
\noindent
\begin{equation}
    \label{eq:expert_policy}
    (\hat{\mathbf{a}}(t),\, \mathbf{h}(t)) = \pi_{\text{PC}}(\mathbf{o}(t),\, \mathbf{h}(t{-}1),\, \tilde{\mathbf{p}}(t)).
\end{equation}
\noindent
Here, $\mathbf{o}(t) \in \mathbb{R}^{3 \times 50}$ denotes the 3D input from 50 HES, $\mathbf{h}(t)$ represents the internal hidden state of the proprietary controller, and $\tilde{\mathbf{p}}(t) \in \mathbb{R}^{6}$ is the 6D reference pose at time $t$. A 6D pose is described as $\mathbf{p} = (x, y, z, \alpha, \beta, \gamma)^T$ in mm and degrees. The proprietary controller $\pi_{\text{PC}}$ leverages a neural network to predict the estimated pose of the mover from $\mathbf{o}(t)$, enabling closed-loop position control based on the estimated and reference poses~\cite{schultePlanarmotorantriebssystemXPlanar2022b}. 
The output $\hat{\mathbf{a}}(t) \in \mathbb{R}^{3 \times 24}$ specifies the actuation commands, where each of the 24 coils is controlled by a triple $\hat{\mathbf{y}}_i = (d_i, q_i, \phi_i)$ corresponding to the $d$-axis current, the $q$-axis current and the electrical phase angle, respectively. To reduce data size, these values are quantized so that each $d_i$ and $q_i$ current is stored as a signed 2-byte integer (\texttt{int16}), and each angle $\phi_i$ is encoded as an unsigned 2-byte integer (\texttt{uint16}) covering one full electrical rotation with wraparound at 65535. In practice, only a subset of coils, typically up to 16 out of 24, are active at each time step, with the remainder effectively disabled (i.e., receiving $d_i = 0$, $q_i = 0$, $\phi_i = 0$). This software architecture enables 6D tracking of reference poses $\tilde{\mathbf{p}}(t)$.

It is important to note that each HES exhibits slightly different characteristics due to manufacturing inaccuracies. These variations are not modeled in this work, as the focus does not lie on generalization across movers and tiles.
Furthermore, HES measurements, typically ranging from $-2000$ to $2000$ increments depending on the mover pose, contain inherent noise with a standard deviation of 4 increments. This leads to a noisy 6D pose estimate predicted by $\pi_{\text{PC}}$, with a standard deviation of approximately $\SI{0.006}{\milli\meter}$ or $0.006^\circ$.

\section{Data Collection}
\label{sec:Data_Collection}

To demonstrate feasibility in this domain, we initially targeted reliable lift-off and sustainable hovering using a behavior-cloned neural controller $\pi_{\text{NC}}$. To accomplish this, we construct a dataset by tracking independent 6D reference trajectories $\tilde{\mathbf{p}}_{n}(t)$ under the control of $\pi_{\text{PC}}$. Each trajectory consists of a lift-off phase (see Fig.~\ref{fig:System}), raising the mover linearly from rest, where the proprietary controller estimates an initial altitude of $z \approx \SI{0.3}{\milli\meter}$, to a uniformly sampled altitude between \SI{2}{\milli\meter} and \SI{4.5}{\milli\meter} over a time window of \SI{1}{\second}, followed by \SI{0.5}{\second} of hovering. The total duration is fixed at $T=\SI{1.5}{\second}$, which is equivalent to $K=6000$ time steps. The remaining five dimensions of the reference pose $(x, y, \alpha, \beta, \gamma)$ remain constant for the entire duration. To adequately cover the workspace, positions $(x,y)$ are sampled from a uniform distribution $\mathcal{U}(-40\text{mm},40\text{mm})$, rotations $(\alpha,\beta)$ from a normal distribution $\mathcal{N}(0^{\circ},0.4^{\circ})$, and yaw rotation $\gamma$ from $\mathcal{U}(-4^\circ,4^\circ)$. Sampling $(x,y,z,\gamma)$ uniformly ensures stable hovering across the full $xyz$ workspace with rotations around the $z$-axis, while sampling rotations $(\alpha,\beta)$ from a normal distribution reflects the typically small orientations around the $x$ and $y$ axes for stable hovering. Furthermore, geometric constraints are enforced on $(z, \alpha, \beta)$ to prevent collisions between the mover and the tile surface. These constraints ensure that the altitude $z$ is sufficient to accommodate the vertical drop of the mover's lowest point caused by rotations $(\alpha, \beta)$ around the $x$ and $y$ axes.

Crucially, standard BC fails to achieve stable levitation due to compounding errors. To overcome this, we introduce a targeted perturbation methodology that addresses covariate shift (see Section~\ref{sec:RelatedWork}) by explicitly eliciting and capturing the expert's recovery behavior. Specifically, we augment each nominal trajectory $\tilde{\mathbf{p}}_n(t)$ with repeated perturbation episodes, yielding perturbed trajectories $\tilde{\mathbf{p}}'_n(t)$ (see Fig.~\ref{fig:System}).
Each perturbation episode consists of two phases: (i) a perturbation phase of duration $\Delta t_{n,\mathrm{pert}}$ during which a linear, continuous deviation from the nominal trajectory in all 6D is induced, and (ii) a correction phase of duration $\Delta t_{n,\mathrm{corr}}$ where the reference pose returns instantaneously, discontinuously, to its nominal value, allowing $\pi_{\text{PC}}$ to re-stabilize the system. The total duration for one perturbation episode is $\Delta t_n = \Delta t_{n,\mathrm{pert}} + \Delta t_{n,\mathrm{corr}}$. Only the start time of the first perturbation episode is uniformly drawn from $\mathrm{U}(0,\, \Delta t_n)$ to ensure dataset-wide variation in perturbation timing. All subsequent episodes follow consecutively until no full perturbation-correction cycle can fit. The perturbation amplitude of reference pose jumps is drawn for each $\tilde{\mathbf{p}}_n(t)$ from a normal distribution $\mathcal{N}(0,\, \xi)$ with $\xi \sim \mathcal{U}(0.1,\, 0.6)$, and subsequently clipped to a maximum of $0.8$ to ensure the proprietary controller remains within valid operational bounds. Both $\Delta t_{n,\mathrm{pert}}$ and $\Delta t_{n,\mathrm{corr}}$ scale with the sampled amplitude to ensure recoverability. This yields the dataset
\begin{equation}
    \mathcal{D}_{\hat{\mathbf{a}} \sim \pi_{\text{PC}}} = \left\{ \tau_n = \left\{ (\mathbf{o}_n(t),\, \tilde{\mathbf{p}}'_{n}(t),\, \hat{\mathbf{a}}_{n}(t)) \right\}_{t=0}^{T} \right\}_{n=0}^{N},
    \label{eq:Dataset}
\end{equation}
\noindent
where $\tau_n$ denotes one lift-off trajectory, $\mathbf{o}_n(t)$ is the observation, $\tilde{\mathbf{p}}'_n(t)$ is the augmented reference pose trajectory, and $\hat{\mathbf{a}}_n(t)$ is the action (coil-level commands) of $\pi_{\text{PC}}$. The resulting dataset contains 11000 trajectories containing approximately \SI{4.6}{\hour} of interaction data and 66 million control steps in total. To prevent unsafe hardware states, the currents of the $d$ and $q$ axes are limited to $\pm8000$ increments based on empirical observations. Although $\pi_{\text{PC}}$ occasionally exceeds this range, hardware constraints are undocumented and, therefore, all values are clipped accordingly during data collection, training, and deployment of the neural controller. 

\section{Neural Control for 6D Magnetic Levitation}
\label{sec:Policy_Training}

Given the augmented dataset defined in Eq.~\eqref{eq:Dataset}, this section addresses the challenge of training a neural controller $\pi_{\text{NC}}$ via BC to imitate the actuation behavior of $\pi_{\text{PC}}$ for 6D MagLev control. The learning problem is characterized by high-dimensional multimodal inputs and outputs, necessitating appropriate encoding, normalization, and a neural architecture capable of modeling temporal dependencies with a strict computation budget of \SI{250}{\micro\second} per time step, which constrains model complexity.

\subsection{Data Representation}

\begin{table}[t]
\vspace{0.95em}
\caption{Worst-case inference time in TwinCAT for 1M executions}
\vspace{-0.5em}
\label{tab:GRUInferenceTimes}
\centering
\begin{tabular}{l|c|c}
\textbf{Model} & $t_{\text{max}} \text{(AVX2)}$ [\si{\micro\second}] & $t_{\text{max}}\text{(scalar)}$ 
[\si{\micro\second}] \\
\hline
GRU(256, 2L) & 137 & 965\\
GRU(256, 4L) & \textbf{242} & 1934\\
GRU(512, 2L) & 478 & 3717\\
\end{tabular}
\vspace{-1.95em}
\end{table}

The neural controller $\pi_{\text{NC}}$ is trained to map HES measurements and reference poses $\mathbf{x}(t) = \text{concat}\!\left(\mathbf{o}(t),\, \tilde{\mathbf{p}}(t) \right)$ to coil current commands $\hat{\mathbf{a}}(t) = \text{concat}\!\left(\mathbf{d}(t), \mathbf{q}(t), \boldsymbol{\phi}(t)\right)$ (see Section~\ref{sec:System_Overview}). HES measurements are standardized using a fixed standard deviation $\sigma_{\text{HES}}$ and zero mean. This fixed scaling ensures that all input channels contribute equally, independent of the mover's workspace. The reference poses are standardized using their empirical mean and standard deviation, computed on the dataset. 
The coil current command $\hat{\mathbf{a}}(t)$ contains three scalars for each coil $(d_{i}(t),\, q_{i}(t),\, \phi_{i}(t))$. As described in Section~\ref{sec:Data_Collection}, the currents for the $d$ and $q$ axes are clamped to $\pm\num{8000}$ in a first step. Next, they are standardized using $\sigma_{\text{dq}} = \num{8000}$ and zero mean.

The electrical phase angle $\phi_{i}(t)$ is periodic. Accordingly, it is first scaled to the range $[0,\, 2\pi]$ and then continuously encoded using a two-dimensional Cartesian representation. Therefore, the neural controller learns the function $\pi_{\text{NC}}: \mathbf{x}(t) \in \mathbb{R}^{156} \rightarrow \hat{\mathbf{y}}(t) \in \mathbb{R}^{96}$ with $\hat{\mathbf{y}}(t) = \text{concat}\!\left(\mathbf{d} \!\left(t \right), \mathbf{q} \!\left(t \right), \cos \!\left(\!2\pi\boldsymbol{\phi}(t)/65535 \!\right), \sin \!\left(\!2\pi\boldsymbol{\phi}(t)/65535 \!\right) \!\right)$.

\subsection{Neural Architecture}

The neural controller $\pi_{\text{NC}}$ must operate under hard real-time constraints, with a strict maximum inference time of \SI{250}{\micro\second} per control step, as dictated by the \SI{4}{\kilo\hertz} cycle rate of the MagLev system. This constraint directly limits the computational complexity of the neural network architecture. The proprietary controller $\pi_{\text{PC}}$, introduced in Section~\ref{sec:System_Overview}, maintains an internal state to capture temporal dependencies over time. To replicate this capability, we adopt a recurrent neural network (RNN) architecture based on gated recurrent units (GRUs), which propagate a hidden state between time steps to retain contextual information from past observations. In contrast to multilayer perceptrons, which require explicit temporal stacking of input features across a fixed time window, increasing both input dimensionality and computational cost proportionally, GRUs maintain constant input size regardless of the time horizon.

\begin{figure}[t]
      \centering
      \includegraphics[width=\columnwidth]{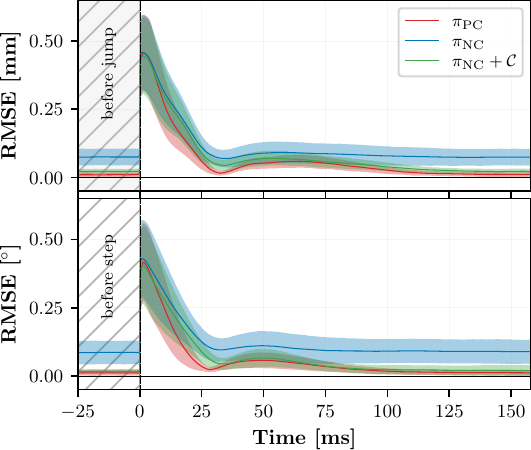}
      \vspace{-1.6em} 
      \caption{Average translational and rotational responses to 100 randomized reference pose jumps, validating bounded-error stability for the proprietary controller $\pi_{PC}$ (red), the neural controller $\pi_{NC}$ (blue), the same neural controller with residual correction network $\pi_{NC} + \mathcal{C}$ (green). Standard deviations are shown as shaded areas.}
      \vspace{-1.5em} 
      \label{fig:StepResponse}
\end{figure}

Therefore, the final architecture comprises four stacked GRU layers, each with 256 hidden units, resulting in a network that exploits the \SI{250}{\micro\second} cycle time (see Table~\ref{tab:GRUInferenceTimes}). 
Ablation studies showed that networks with reduced depth lack sufficient generalization for robust mover levitation. 
At each control step, $\pi_{\text{NC}}$ operates autoregressively by consuming the input $\mathbf{x}(t)$, updating its internal state, and producing the actuation command $\mathbf{a}(t)$.

\subsection{Training Procedure}
\label{sec:Training}

Training is performed offline via supervised BC using \texttt{float32} precision, Adam optimizer with $lr=1 \times 10^{-4}$, and a batch size of 256. The model is trained on $10000$ and validated on $1000$ randomly shuffled lift-off trajectories. 

To address covariate shift and improve generalization beyond the dataset's observation distribution, $\ell_2$ regularization is applied using three weight decay coefficients $\lambda \in \{6.25 \times 10^{-8},\ 2.5 \times 10^{-7},\ 1 \times 10^{-6}\}$. Among these, only $\lambda = 2.5 \times 10^{-7}$ yields a neural controller that consistently tracks unseen reference trajectories without loss of control.

An adaptive learning rate schedule reduces the learning rate based on validation loss, and training is terminated after $10$ consecutive epochs without improvement. Training completes within two days on an NVIDIA RTX 6000 Ada GPU. The neural controller is trained on input sequences of $50$ time steps, producing predictions at each step. Loss is applied only to the final 10 steps to allow hidden state warmup. Gradients are backpropagated through the full sequence to enable learning of long-range dependencies.

The loss function is mean squared error (MSE), with a zero-masking strategy to avoid penalizing outputs for inactive actuators. Specifically, if both the dataset's and the predicted $d_j$ and $q_j$ currents for coil $j$ are below a threshold of five current increments, all loss terms for $d_j$, $q_j$, and $\phi_j$ at that time step are excluded. This is particularly important for $\phi_j$, whose value is irrelevant when no current is applied and therefore no magnetic field is generated.
 \subsection{Deployment}

As described in Section~\ref{sec:System_Overview}, TwinCAT tasks operate in a real-time kernel without access to external libraries. Consequently, the GRU-based inference of $\pi_{\text{NC}}$ must be implemented as a fully custom C++ module to meet hard real-time constraints. Model complexity is therefore strictly coupled to the real-time limit of \SI{250}{\micro\second} per control cycle. To support more expressive models within this bound, the implementation must be highly optimized for low-latency execution. 
Our implementation uses a custom binary serialization format compatible with PyTorch and TwinCAT, and is optimized through heavy use of Advanced Vector Extensions~2~(AVX2) intrinsics for vectorized matrix operations and approximated activation functions. Because hard real-time faults occur if a single cycle time is exceeded, Table~\ref{tab:GRUInferenceTimes} reports the critical worst-case execution time for GRU-based architectures. For the tested neural architectures, the AVX2-optimized implementation achieves an $8\times$ speedup over a naive scalar baseline.

\section{Experimental Evaluation}
\label{sec:Experimental_Evaluation}

\begin{figure}[tpb]
      \centering
      \includegraphics[width=\columnwidth]{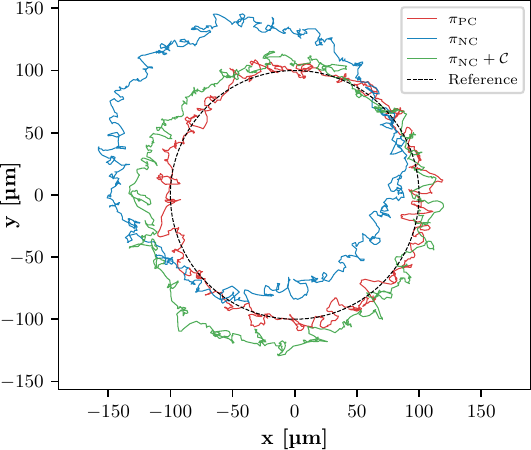}
      \vspace{-1.6em} 
      \caption{Accuracy measurements for a circular trajectory tracking experiment regarding the proprietary controller $\pi_{PC}$ (red), the neural controller $\pi_{NC}$ (blue), the same neural controller with residual correction network $\pi_{NC} + \mathcal{C}$ (green), and the reference circle with a radius of $\SI{100}{\micro\meter}$ (dashed black).}
      \vspace{-1.5em}
      \label{fig:CircleTrajectory}
\end{figure}

\begin{figure}[t]
      \centering
      \scalebox{1}[1.007]{\includegraphics[width=\columnwidth]{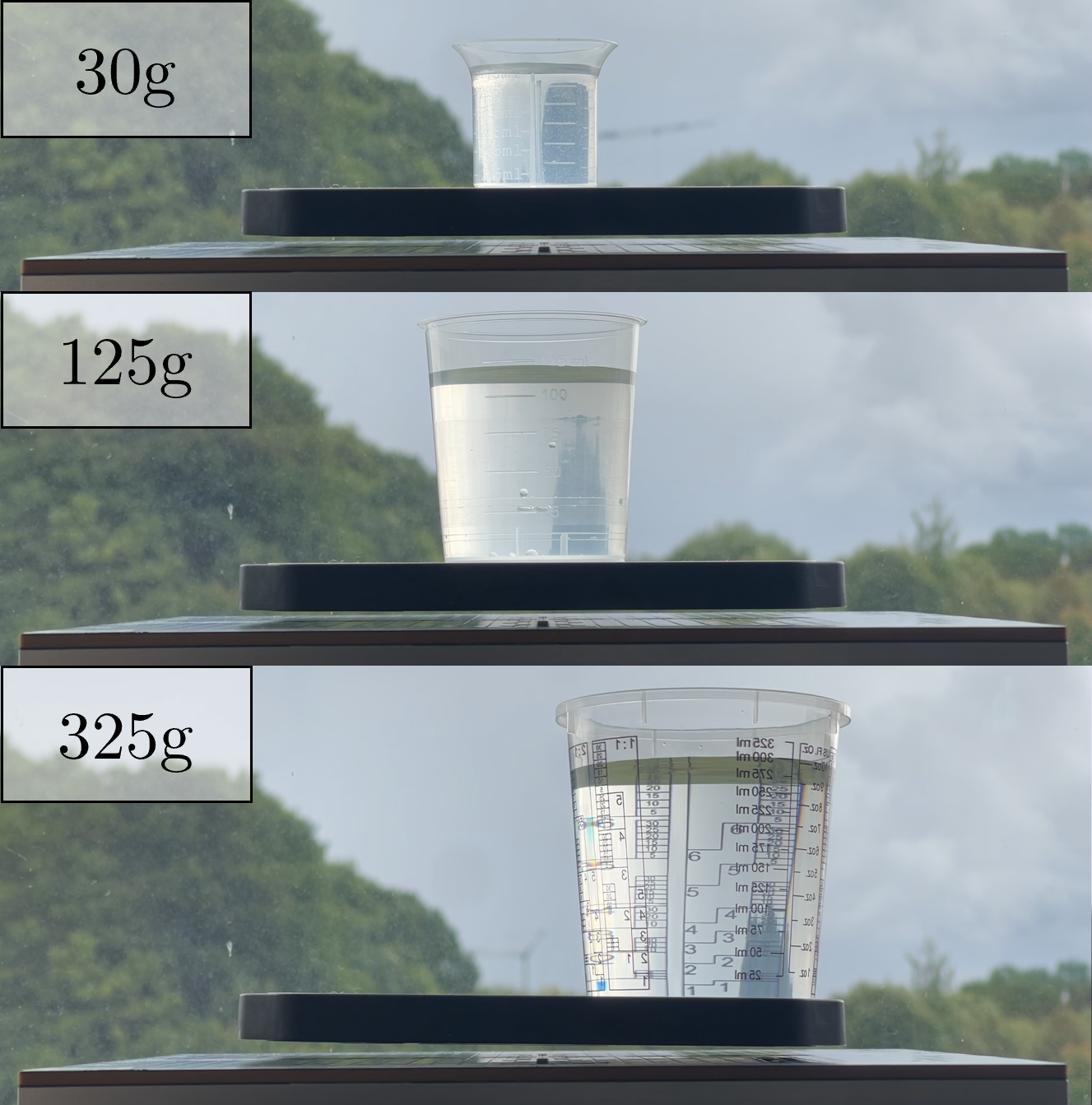}}
      \vspace{-1.6em} 
      \caption{The neural controller $\pi_{\text{NC}}$ remains stable while hovering despite carrying payloads of varying masses and mass distributions.}
      \vspace{-1.5em} 
      \label{fig:Weights}
\end{figure}

\begin{figure}[th]
      \centering
      \includegraphics[width=\columnwidth]{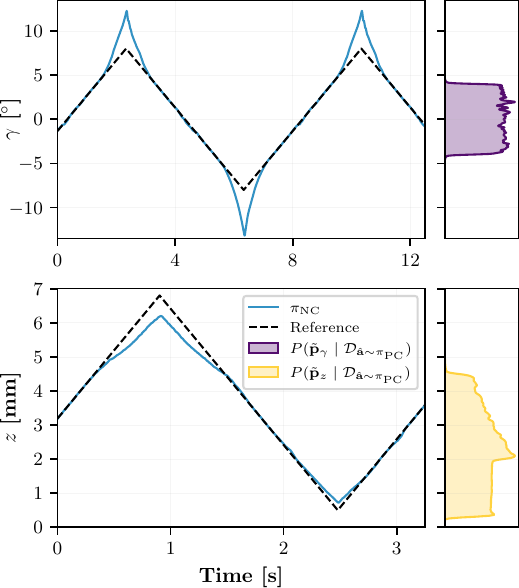}
      \vspace{-1.6em} 
      \caption{The neural controller $\pi_{\text{NC}}$ (blue) tracks reference trajectories (dashed black) beyond the training distribution in $\gamma$ (purple shaded area) and $z$ (yellow shaded area). }
      \vspace{-1.5em} 
      \label{fig:extrapolation}
\end{figure}

The evaluations rely on the pose estimations of the proprietary controller, serving as the most accurate available proxy for actual mover poses. During initial tests, we observed systematic deviations between the reference poses and estimated poses for $\pi_{NC}$ (up to $0.5\,\text{mm}$ or $0.5^\circ$ per axis). These deviations reflect the inherent covariate shift problem of BC, as the neural controller is trained without online system interaction. 

To address these systematic deviations, we introduce a residual correction network $\mathcal{C}$, realized by a multilayer perceptron (MLP).
Here, the system is locally linearized around the reference pose to correct for systematic pose deviations. During an initial data collection phase, a set of reference poses $\tilde{\mathbf{p}}(t)$ and the corresponding estimated poses $\mathbf{p}(t)$ are recorded. This dataset is then used to train the MLP to learn a mapping from the reference pose to the expected systematic error. Formally, the original reference pose $\tilde{\mathbf{p}}(t)$ is mapped to a calibrated version $\tilde{\mathbf{p}}'(t)$ by
$\tilde{\mathbf{p}}'(t) = \tilde{\mathbf{p}}(t) + \mathcal{C}(\tilde{\mathbf{p}}(t))$, where $\mathcal{C}(\tilde{\mathbf{p}}(t)) = \tilde{\mathbf{p}}(t) - \mathbf{p}(t)$, and $\mathbf{p}(t)$ denotes the estimated pose.

The resulting residual-corrected neural controller, denoted as~$\pi_{\text{NC}}+\mathcal{C}$, utilizes the calibrated reference pose as input and demonstrates reduced deviations between $\tilde{\mathbf{p}}(t)$ and $\mathbf{p}(t)$.
To validate $\pi_\text{NC}$ and $\pi_\text{NC} + \mathcal{C}$, we empirically evaluate stability and generalization beyond the original training trajectories. 

\subsection{Stability Analysis}
\label{sec:Stability}

Although formal stability guarantees represent a promising direction for future work, we provide here an empirical bounded-error analysis utilizing the real system. 
We evaluate sustained control under both discontinuous and continuous disturbances. In addition, Section~\ref{sec:generalization} presents successful tracking of a randomly generated continuous trajectory without loss of control.

\begin{table}[b]
\vspace{-1.6em}
\caption{Root mean squared error between reference pose and estimated pose for a random trajectory}
\vspace{-0.5em}
\label{tab:rmse_randomtraj}
\centering
\begin{tabular}{l|cccccc}
\textbf{Method} & \textbf{x} & \textbf{y} & \textbf{z} & $\boldsymbol{\alpha}$ & $\boldsymbol{\beta}$ & $\boldsymbol{\gamma}$ \\
\hline
$\pi_{\text{PC}}$ & 0.107 & 0.096 & \textbf{0.012} & \textbf{0.018} & \textbf{0.013} & \textbf{0.020} \\
$\pi_{\text{NC}}$ & 0.077 & 0.119 & 0.091 & 0.073 & 0.082 & 0.134 \\
$\pi_{\text{NC}}+\mathcal{C}$ & \textbf{0.069} & \textbf{0.077} & 0.021 & 0.023 & 0.024 & 0.035 \\
\end{tabular}
\end{table}

To evaluate sustained control under discontinuous disturbances, we apply random reference pose steps to test the recoverability of $\pi_{\text{NC}}$ and $\pi_{\text{NC}} + \mathcal{C}$. Each trial introduces instantaneous reference pose jumps across all six degrees of freedom, with magnitudes up to \SI{0.8}{\milli\meter} or \SI{0.8}{^\circ} per axis. The initial reference pose is sampled uniformly within the 6D workspace. Fig.~\ref{fig:StepResponse} shows the root mean squared error (RMSE) between the estimated pose and the reference pose before and after the step. 
The results demonstrate empirical bounded-error stability for all controllers. 
The residual-corrected neural controller $\pi_{\text{NC}}+\mathcal{C}$ achieves improved absolute accuracy, despite operating without explicit closed-loop position control.

\begin{figure*}[tpb]
    \centering
    \includegraphics[width=\textwidth]{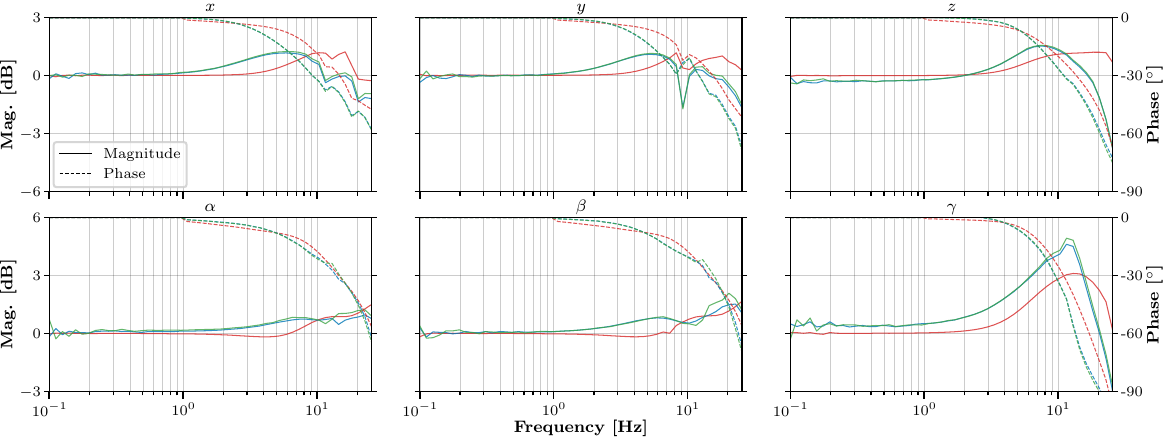}
    \vspace{-1.6em} 
    \caption{Bode plots illustrating frequency responses across translational ($x, y, z$) and rotational ($\alpha, \beta, \gamma$) axes for perturbations of \SI{0.5}{\milli\meter} or \SI{0.5}{^\circ} for \mbox{$\pi_\text{PC}$ (red)}, $\pi_\text{NC}$ (blue), and $\pi_\text{NC}+\mathcal{C}$ (green).}
    \vspace{-1.5em} 
      \label{fig:Bode_Plots}
\end{figure*}

To evaluate sustained control under continuous disturbances, we performed a frequency-based robustness test using single-axis sinusoidal excitation with fixed amplitude (\SI{0.5}{\milli\meter} or \SI{0.5}{^\circ}) applied across frequencies from \SI{0.1}{\hertz} to \SI{25.6}{\hertz}. Fig.~\ref{fig:Bode_Plots} shows the resulting Bode plots for all controllers. 
Analysis of the frequency response reveals that the neural controllers, $\pi_{\text{NC}}$ and $\pi_{\text{NC}}+\mathcal{C}$, effectively mimic the proprietary controller $\pi_{\text{PC}}$, but with slightly reduced overall robustness. At low frequencies ($<$\SI{1}{\hertz}), both neural and proprietary controllers show minimal phase lag. As frequency increases, performance diverges. While $\pi_{\text{NC}}+\mathcal{C}$ shows transient advantages (e.g., lower phase lag in $z$ up to \SI{5}{\hertz} and better high-frequency rejection in $x$ and $y$ between \SI{10}{}-\SI{20}{\hertz}), the proprietary controller $\pi_{\text{PC}}$ generally exhibits superior performance, characterized by lower phase lag and less magnitude amplification, particularly in the $\gamma$-axis. However, despite these quantitative differences, both $\pi_{\text{NC}}$ and $\pi_{\text{NC}}+\mathcal{C}$ maintained stability across the full \SI{0.1}{}-\SI{25.6}{\hertz} spectrum on all axes. In addition, $\pi_{\text{NC}}$ and $\pi_{\text{NC}}+\mathcal{C}$ replicate characteristics of $\pi_\text{PC}$, such as specific magnitude and phase dips along the $y$-axis, despite never being trained on harmonic excitations. The results of the Bode plots confirm their capacity to handle continuous disturbances without failure, validating the robustness of the neural controller.

\subsection{Generalization Analysis} 
\label{sec:generalization}

In this section, we evaluate all controllers on trajectories that go beyond the training distribution. Any nonlinear trajectory without discontinuous reference pose steps demands substantial generalization capability, as the distribution of inputs (reference poses and Hall sensor measurements) is not represented in the training data. Fig.~\ref{fig:CircleTrajectory} illustrates the tracking of a circular trajectory (\SI{100}{\micro\meter} radius at \SI{4}{\hertz}). Although $\pi_\text{PC}$ remains best at tracking the trajectory in the $xy$-plane, all controllers remain stable throughout the entire trajectory, showcasing advanced generalization. The residual-corrected neural controller $\pi_{\text{NC}}+\mathcal{C}$ clearly surpasses $\pi_{\text{NC}}$, demonstrating that residual correction significantly improves absolute accuracy. 

An even more rigorous generalization test is performed on a randomized continuous trajectory across the 6D workspace with varying dynamics. As can be seen in Table~\ref{tab:rmse_randomtraj}, $\pi_{\text{NC}}+\mathcal{C}$ achieves superior pose tracking accuracy in $x$ and $y$, even compared to $\pi_{\text{PC}}$. Since $\pi_\text{NC}$ already has superior pose tracking accuracy in $x$, this indicates that the generalization capability of the neural controller to unseen state distributions leads to improved tracking accuracy compared to the proprietary controller. This demonstrates not only strong generalization to unseen random trajectories without loss of control, but even partial improvements over $\pi_\text{PC}$. 

In addition to trajectory evaluations, we also examine generalization under distributional shifts in system dynamics. As shown in Fig.~\ref{fig:Weights}, despite no samples of any weight-carrying in the dataset, the neural controller remains stably levitated under payloads of \SI{30}{g}, \SI{125}{g}, and up to \SI{325}{g}, near its upper limit. This demonstrates extrapolation to unseen physical conditions. To further assess extrapolation, out-of-distribution poses were commanded beyond the ranges seen during training (see Fig.~\ref{fig:extrapolation}). The training data for $\pi_{\text{NC}}$ cover yaw rotations up to \SI{4}{^\circ} with maximum augmentations clipped at \SI{0.8}{^\circ} (total \SI{4.8}{^\circ}), and altitudes up to \SI{4.5}{\milli\meter} with maximum augmentations of \SI{0.8}{\milli\meter} (total \SI{5.3}{\milli\meter}). In contrast, the neural controller can achieve altitudes of up to \SI{6.2}{\milli\meter} at a reference altitude of \SI{6.8}{\milli\meter} and rotations of up to \SI{12.8}{^\circ} at a reference rotation of \SI{8}{^\circ}. The systematic deviations between the reference pose and the estimated pose prove to be significantly greater in poses beyond the training distribution. However, the neural controller maintains stability even at rotation angles $2.5\times$ larger than those covered by demonstrations of the proprietary controller. In fact, the proprietary controller supports unconstrained positioning only for rotations in $\gamma$ of up to \SI{5}{^\circ} and altitudes in $z$ of up to \SI{5}{\milli\meter}~\cite{Beckhoff_XPlanar_2023_v1_3}.

\section{Discussion \& Future Directions}
\label{sec:discussion}

This work demonstrates that an E2E neural controller trained solely on augmented lift-off trajectories via BC can control an industrial-grade 6D MagLev system accurately and robustly. The neural controller generalizes well beyond the training data, robustly following random trajectories despite being trained only on augmented lift-off trajectories. In addition to strong generalization, the controller also extrapolates to altered dynamics, such as untrained weight loads and operation outside the pose ranges covered in the dataset. This level of robustness, generalization, and extrapolation demonstrates the potential of E2E neural control in high-frequency, high-dimensional, inherently unstable, and complex industrial-grade systems.
A primary challenge of offline BC is covariate shift, which manifests in our experiments as systematic deviations between the reference and estimated mover poses. Therefore, a residual correction network further improves absolute accuracy, enabling $\pi_{\text{NC}}+\mathcal{C}$ to surpass the proprietary controller $\pi_{\text{PC}}$ in certain scenarios. The introduced residual correction network operates outside the neural control and is trained post hoc, which could be seen as a contradiction of a fully E2E approach. However, its success provides a critical insight: the control errors induced by covariate shift are highly structured and therefore learnable.
Consequently, we argue that this strongly motivates further interaction-based learning methods like reinforcement learning or model predictive control that can intrinsically eliminate such errors. Here, the presented behavior-cloned neural controller can serve as a robust and safe baseline. Therefore, the success of BC should not be understood as wholly outperforming expert-engineered control, but as a strong demonstration of the practicality, robustness, and potential of neural control in such systems. 
Therefore, the challenges of BC, such as covariate shift and the reliance on a closed-source expert, strongly motivate the development of open-source, white-box baselines for rigorous comparative benchmarking.
Another promising research direction involves multi-tile and multi-mover systems, additionally accounting for manufacturing variations across different tiles and movers.


\bibliographystyle{IEEEtran}
\bibliography{IEEEabrv,references}

\end{document}